# Evolution of magnetic phase in two dimensional van der Waals Mn$_{1-x}$Ni$_x$PS$_3$ single crystals

Ziye Lu[1], Xinyu Yang[1], Lin Huang[2], Xiyu Chen[3], Meifeng Liu[3], Jin Peng[1*], Shuai Dong[1*], and Jun-Ming Liu[2]

1. School of Physics, Southeast University, Nanjing 211189, China

2. Laboratory of Solid State Microstructures, Nanjing University, Nanjing 210093, China

3. Institute for Advanced Materials, Hubei Normal University, Huangshi 435002, China

**Abstract**

Metal thio(seleno)phosphates $M$PX$_3$ have attracted considerable attentions with wide spanned band gaps and rich magnetic properties. In this series, two neighboring members MnPS$_3$ and NiPS$_3$ differ in magnetic atoms, magnetic easy axes, spin anisotropy, as well as nearest-neighbor magnetic interactions. The competition between these components may cause intriguing physical phenomena. In this article, the evolution of magnetism of Mn$_{1-x}$Ni$_x$PS$_3$ series is reported. Despite the incompatible antiferromagnetic orders of two end members, the antiferromagnetism persists as the ground state in the whole substitution region. The magnetic ordering temperature $T_N$ show nonmonotonic V-shape behavior, and the reentrant spin glass phase at $x$=0.5 is observed. In addition, abnormal bifurcation of $T_N$ occurs at $x$=0.75, which may be due to the temperature-dependent spin reorientation or phase separation. The evolution of magnetism is further confirmed semi-quantitatively by our density functional theory calculations. Our study indicates that exotic magnetism can be intrigued when multi-degrees of freedom are involved in these low-dimensional systems, which call for more in-depth microscopic studies in future.

*E-mails: jpeng@seu.edu.cn; sdong@seu.edu.cn

## 1. Introduction

Two-dimensional (2D) van der Waals (vdW) materials are those systems in which monolayer atoms or several atomic layers are stacked by vdW force. They have attracted considerable interests since the successful exfoliation of graphene [1]. Since then, huge amounts of 2D vdW materials have been synthesized, including hexagonal boron nitride ($h$-BN) [2], transition metal dichalcogenides [3, 4], graphitic carbon nitride ($g$-$C_3N_4$) [5] and so on. Most 2D vdW materials in early years are non-magnetic. However, 2D vdW magnets are highly desired, which can host exotic quantum phases and play as material base for quantum devices. In contrast to another quasi-2D magnetic materials, such as magnetic thin films, 2D vdW magnetic materials can be superior in performances, such as thinner, free of substrates, and less corroded from surfaces and defects. These specialties make them mechanically flexible and easy for chemical functionalization.

In addition, 2D vdW magnets can be a good test field of some early-established theories. For example, Mermin and Wagner once predicted "absence of ferromagnetism or antiferromagnetism in one-dimensional (1D) or 2D isotropic Heisenberg models at any non-zero temperature" [6]. This theorem implies that long-range magnetic order in 2D systems with continuous rotational symmetry is excluded in nonzero temperature due to thermal fluctuations. Even though, it can be stabilized by gapping the low-energy modes through the introduction of anisotropy [7-9]. Recently, long-range ferromagnetic order was realized in 2D monolayers or few-layers exfoliated from vdW materials, such as $CrI_3$ [10], $Cr_2Ge_2Te_6$ [8], and $Fe_3GeTe_2$ [11]. These observations not only change the physical understanding of two-dimensional magnets, but also shed light on potential applications of nonvolatile, low-power consumption spintronic devices [12, 13]. Since then, the researches of vdW 2D magnets grow rapidly [14-20].

The family of $M$PS$_3$ compounds ($M$=3$d$ transition metal) offers a series of isostructural vdW layered compounds with plenty antiferromagnetism and spin anisotropy [21-23]. Although antiferromagnets display zero net magnetization, they can response to external perturbations and display ultrafast spin dynamics, for which the so-called antiferromagnetic spintronics have emerged in recent years. In addition, the wide spanned band gap of 1.2-3.5 eV of $M$PS$_3$ also make them promising for applications in optoelectronics and catalysis [24].

The $M$PS$_3$ series adopt a monoclinic structure with space group $C2/m$. Transition-metal ions $M$ are caged in trigonally distorted sulfur octahedra, form honeycomb networks in the $ab$ planes, and carry local magnetic moments [25], as shown in figure 1(a). A pair of phosphorous atoms locates at the center of each honeycomb ring. Layers stack along the trigonal axis with vdW force, as shown in figure 1(b). This layered structure can be mechanically exfoliated to few, even single layers.

With different transition metal ions, $M$PS$_3$ possess diverse magnetic properties. The competitions between the direct $M$-$M$ exchange and indirect superexchange mediated through S atoms within each layer, as well as the interlayer exchange, decide the magnetic ground states and ordering temperatures [26]. The Néel temperature ($T_N$) varies with transition metal $M$, from 78 K for MnPS$_3$, to 118 K, and 155 K for FePS$_3$, and NiPS$_3$, respectively [6, 27-31]. The magnetic structure of MnPS$_3$ is shown in figure 1(c). Below $T_N$, Mn ions are antiferromagnetically coupled with three nearest neighbors within the basal plane, ferromagnetically coupled with its interlayer nearest neighbor, resulting a two-dimensional Néel type antiferromagnetism. The direction of spins are 8° canting from the $c$-axis [32, 33]. Whereas in NiPS$_3$, Ni ions are antiferromagnetically coupled with one nearest neighbor, ferromagnetically coupled with other two nearest neighbors, forming the zigzag antiferromagnetic state, as show in figure 1(d) [26]. The magnetic moments of Ni lie predominantly along the $a$-axis with a small component along the $c$-axis as shown in figure 1(e). FePS$_3$ also owns the zigzag type antiferromagnetism, while the moments align along the out-of-plane direction.

Another fundamental parameter in collinear magnets is the magnetic anisotropy, coming from variety of sources (e.g. magnetocrystalline anisotropy, stress anisotropy, spin-orbit coupling, and shape anisotropy). In 2D magnets, such anisotropy is crucial in establishing long-range correlation since that it can gap the low-energy modes [34]. For example, in FePS$_3$, trigonal distortion combining with the spin-orbit coupling yields a large single-axis magnetic anisotropy [35], thus it can be described by the Ising model roughly. While, in NiPS$_3$, the anisotropy is rather weaker, it is better be described by anisotropic Heisenberg model or XY model [26]. In MnPS$_3$, the intrinsic anisotropy almost diminishes, thus it can be modeled using a Heisenberg Hamiltonian [27].

Given these distinct magnetic order, spin direction, and magnetic anisotropy, exotic quantum phases are expected to be revealed in their solutions. For example, spin glass behavior was found in Mn-substituted FePS$_3$ [36-38]. A slow relaxation of magnetic state was observed in Ni doped FePS$_3$ [39].

In this work, we systematically studied the magnetic properties of 2D vdW layered Mn$_{1-x}$Ni$_x$PS$_3$ series. Possible temperature-dependent spin reorientation (or phase separation) in Mn$_{0.25}$Ni$_{0.75}$PS$_3$ and glassy behavior in Mn$_{0.5}$Ni$_{0.5}$PS$_3$ are reported, revealing the complex competition between magnetic exchange interactions and anisotropy.

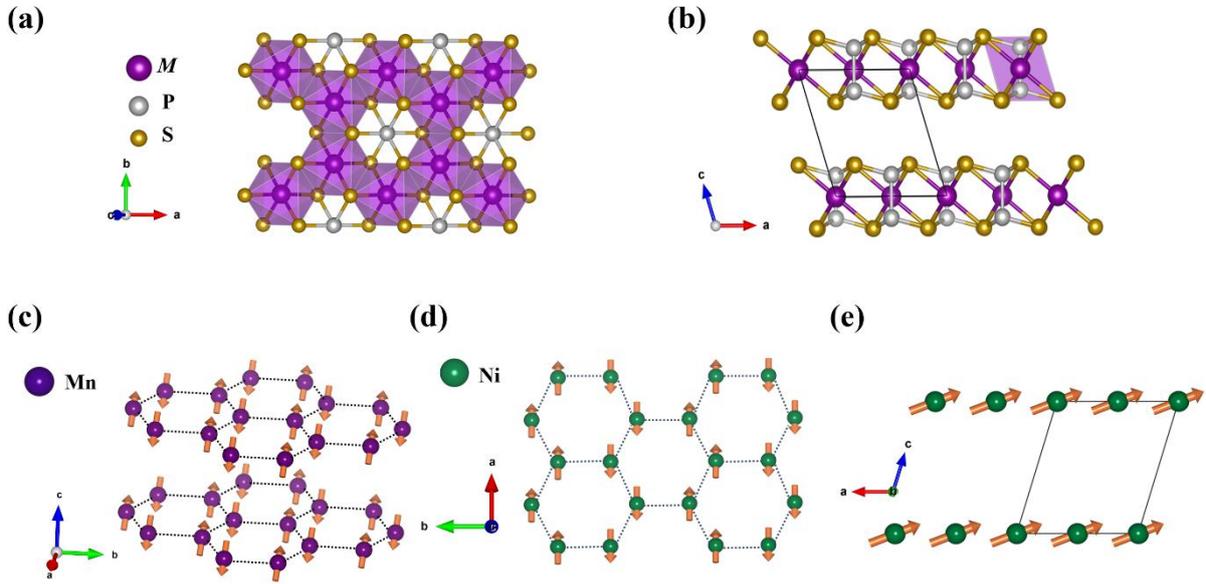

**Figure 1.** Structure of $M$PS$_3$ ($M$ = Ni/Mn). (a) Top view and (b) side view of the crystal structure. (c) Magnetic structure of MnPS$_3$: a Néel type antiferromagnetism with spins pointing close to the $c$-axis. (d) Top view and (e) side view of magnetic structure of NiPS$_3$: a zigzag type antiferromagnetism with spins lying close to the $ab$-plane.

## 2. Details of experimental and calculation methods

### 2.1. Synthesis of Layered Metal Thiophosphates.

Single crystals of Mn$_{1-x}$Ni$_x$PS$_3$ series were synthesized using the chemical vapor transport technique, as shown in figure 2(a). In an argon-filled glove box, manganese powder (99.95%), nickel powder (99.95%), phosphorus chunks (99.999%), and sulfur powder (99.95%) with stoichiometric molar ratio of $Mn_{1-x}Ni_xPS_3$ were mixed and loaded into quartz tubes (20-mm-outside-diameter and 17-mm-inside-diameter) with transport agent iodine. The tubes were sealed with an oxy-hydrogen gas torch under a pressure of $10^{-3}$ Pa. The sealed tubes were heated in a two-zone furnace. The hot and cold zone were heated to 700 °C and 650 °C respectively with a ramp rate of 1 °C per minute and were hold for 7 days, followed by furnace cooling. The grown crystals formed shiny flakes with typical dimensions of 5 mm×5 mm×60 μm, as shown in figure 2(b).

## 2.2. Characterization and magnetic measurements.

To identify the crystal structures of samples, X-ray diffraction (XRD) patterns were measured by Rigaku Smartlab3 (40 kV, 30 mA) in Bragg-Brentano reflection geometry with Cu $K\alpha$ radiation ($\lambda$=1.5406 Å). The electron microscopic images were captured by scanning electron microscope (SEM). The same SEM device allowed for energy dispersive x-ray spectroscopy (EDXs) measurements (Oxford INCA Energy) exam the chemical distribution. A thin film of gold was evaporated on the surface of samples to further improve the sample conductivity. Magnetization measurements were carried using Superconducting Quantum Interference Device (Quantum Design). Temperature dependence of susceptibilities were measured in an applied magnetic field of $\mu_0H$=0.5 T, in both the zero field cooling (ZFC) and field cooling (FC) sequences. Field dependent magnetization were carried under different temperatures. Thermoremanent magnetization were measured after ZFC to 20 K and 2 K, then set the magnetic field to 5 T and lasted for 300 second, and then removed the magnetic field.

## 2.3. First-principles calculations.

First-principles calculations were performed based on density functional theory (DFT) as implemented in the Vienna *ab initio* simulation package (VASP) [40]. For the exchange-correlation functional, the PBE parametrization of the generalized gradient approximation (GGA) was used [41], and the Hubbard $U$ was applied using the Dudarev parametrization [42]. According to our test, $U_{eff}$=3 eV for both Mn's and Ni's 3$d$ orbitals are used, which can lead to best lattice constants comparing with the experimental ones (see figure S1 in Supplementary Materials). The energy cutoff is set as 400 eV and the $k$-point grid is 5×3×5 for the bulks. The convergence criterion for the energy is $10^{-6}$ eV for self-consistent iteration, and the Hellman-Feynman force was < 0.01 eV/Å during the structural relaxation. And the van der Waals correction DFT-D3 method was adopted [43, 44].

## 3. Results and discussions

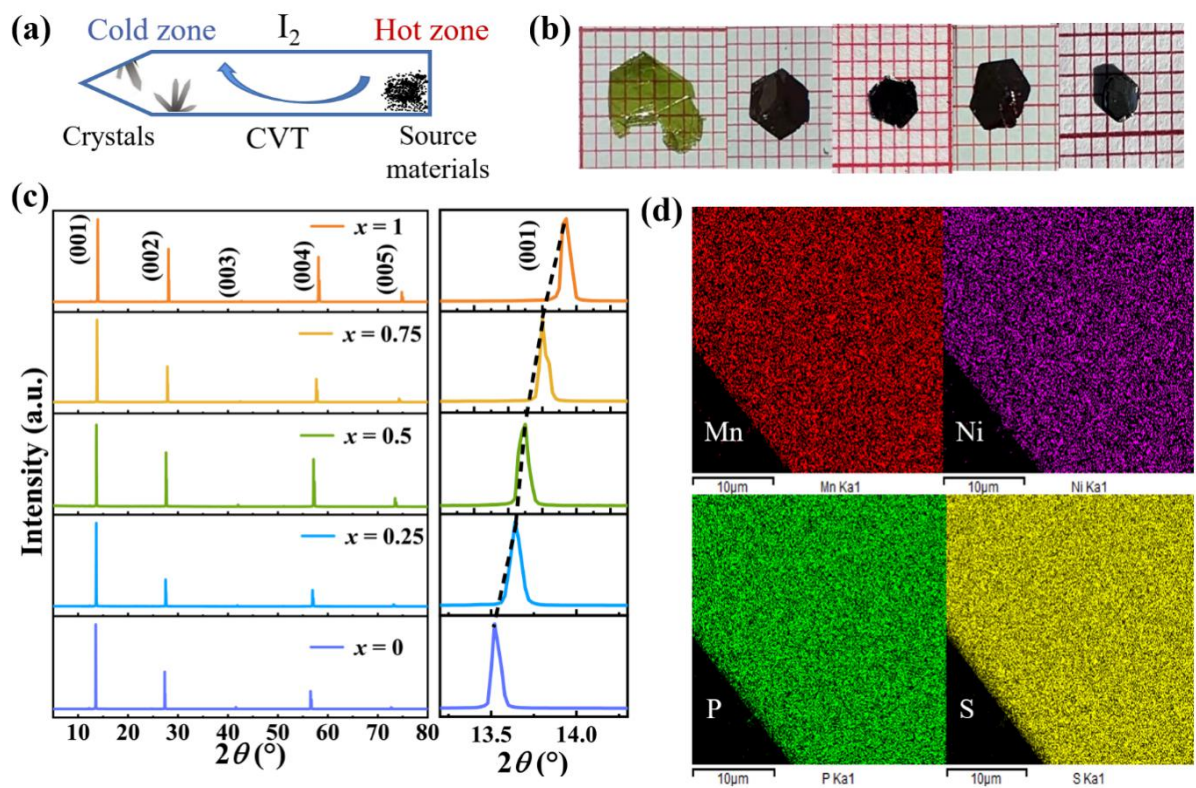

**Figure 2.** (a) Schematic of two zone chemical vapor transport set-up. (b) Pictures of as-grown crystals from left to right: $MnPS_3$, $Mn_{0.75}Ni_{0.25}PS_3$, $Mn_{0.5}Ni_{0.5}PS_3$, $Mn_{0.25}Ni_{0.75}PS_3$, and $NiPS_3$; a red square in the background corresponds to 1 mm × 1 mm for scale. (c) XRD patterns for single crystalline $Mn_{1-x}Ni_xPS_3$ series showing the (00$L$) reflections. Right panel shows the movement of (001) diffraction peaks. (d) EDXs elemental mappings of $Mn_{0.5}Ni_{0.5}PS_3$.

The phases of $Mn_{1-x}Ni_xPS_3$ series were first characterized by XRD. As shown in figure 2(c), the (00$L$) diffraction peaks shift to high angle gradually with increasing Ni content $x$, which is in line with the larger ionic radius of $Mn^{2+}$ compared to $Ni^{2+}$. The successful substitution of Ni were also verified by EDXs measurements. Detailed results are shown in Table S1 of Supplementary Material. To get information about the elements distribution, an element mapping were carried out on the crystal surface as shown in figure 2(d). An even distribution of Mn, Ni, P, and S was observed.

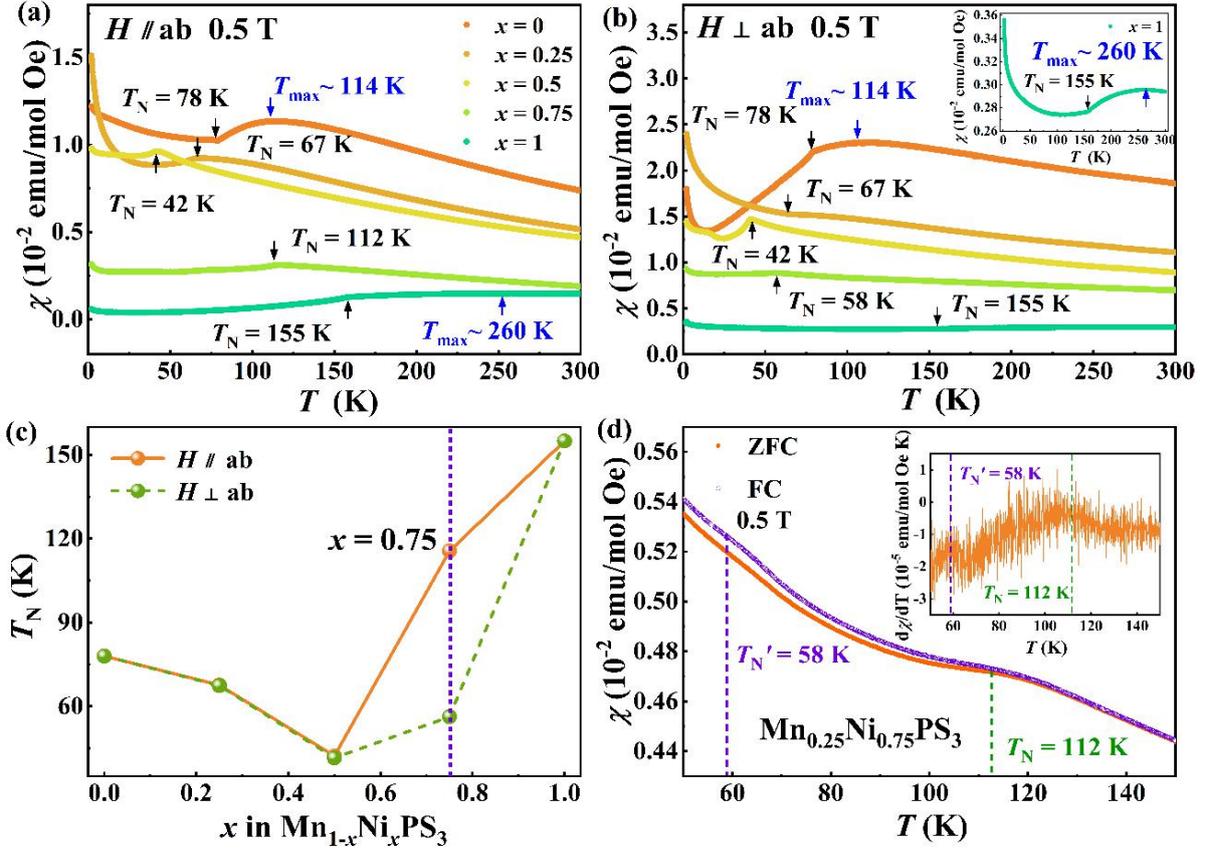

**Figure 3.** The magnetic susceptibilities of $Mn_{1-x}Ni_xPS_3$ single crystals under magnetic fields of 0.5 T applied parallel and perpendicular to the *ab* plane as a function of temperature. (a) $H \parallel ab$; (b) $H \perp ab$. Inset of (b): magnetic susceptibility of $NiPS_3$ under an out-of-plane magnetic field of 0.5 T. (c) Antiferromagnetic ordering temperature $T_N$ as a function of Ni content $x$ in $Mn_{1-x}Ni_xPS_3$ for magnetic fields applied along and perpendicular to the *ab* plane. (d) The FC and ZFC magnetic susceptibilities of polycrystalline $Mn_{0.25}Ni_{0.75}PS_3$ under a magnetic field of 0.5 T as a function of temperature. Inset in (d): $d\chi/dT$.

As shown in figures 3(a) and 3(b), under both in-plane ($H \parallel ab$) and out-of-plane ($H \perp ab$) magnetic fields, the magnetic susceptibilities exhibit a systematic decrease with increasing Ni content $x$, which can be ascribed to a larger magnetic moment of $Mn^{2+}$ ($S=5/2$, ~5 $\mu_B$) compared to $Ni^{2+}$ ($S=1$, ~2 $\mu_B$). In addition, the shapes of susceptibility curves evolve upon the substitution. For two end members $MnPS_3$ and $NiPS_3$, broad maximums were observed for both $H \parallel ab$ and $H \perp ab$, as indicated by blue arrows in figures 3(a) and 3(b). This broad maximum is assumed to be related to the strong short-range spin correlation above the long-range ordering temperature in low-dimensional systems [45]. Below $T_{max}$, an inflection point which reflects the onset of long-range order was defined as the real $T_N$. For three doped members, the maximums are not observed, which is attributed to the quenching disorder introduced by substitution.

The magnetic ordering temperatures $T_N$'s, defined by the inflection points, are 78 K and 155 K for MnPS$_3$ and NiPS$_3$ respectively, which are consistent with the previous report [31]. With increasing concentration of Ni, it shows non-monotonic change. As summarized in figure 3(c), $T_N$ first decreases to the lowest point at $x$=0.5, and then increases with further increasement of Ni. This non-monotonic V-shape evolution of $T_N$ is due to the incompatible magnetic orders of MnPS$_3$ and NiPS$_3$, similar to the situation in Mn$_{1-x}$Fe$_x$PSe$_3$ [46]. In contrast, $T_N$ increases gradually over the full substitution range from FePS$_3$ to NiPS$_3$ [47], both of which own the zigzag type antiferromagnetic order.

Below $T_N$, there is also an evolution on the ordered moments' directions. For MnPS$_3$, susceptibility curve drops more quickly for $H \perp ab$ below $T_N$, indicating spins order predominately out-of-plane. For NiPS$_3$, susceptibility curve drops more quickly for $H \parallel ab$ below $T_N$, indicating spins order predominately in-plane. For the half doped sample, susceptibility curve still drop more quickly for $H \perp ab$ below $T_N$. It is quite possible that spin orders predominately out-of-plane, similar to MnPS$_3$. For $x$=0.75, situation becomes complicated.

A noteworthy feature in figure 3(c) is that there are two reflection points (112 K for $H \parallel ab$ and 58 K for $H \perp ab$, defined as $T_N$ and $T'_N$ respectively) in the $x$=0.75 sample, which was also evidenced but not discussed in a previous study [48]. To further confirm this bifurcation phenomenon, we measured the magnetic susceptibility of powdered Mn$_{0.25}$Ni$_{0.75}$PS$_3$ single crystals. As shown in figure 3(d), two anomalies were observed at ~112 K and ~58 K. This feature can be also evidenced in the derivative d$\chi$/d$T$, as shown in the inset of figure 3(d).

We further measured the isothermal magnetization of this $x$=0.75 compound. As shown in figures 4(a) and 4(b), for $H \parallel ab$ and $H \perp ab$, the magnetization shows antiferromagnetic like behaviors. However, the magnitudes of magnetization along different directions show distinct trends with temperature. The net magnetic moment curves at 7 T for both $H \parallel ab$ and $H \perp ab$ are plotted in figure 4(d) as a function of temperature. In the high temperature region, both values increase with decreasing temperature, as expected for the suppression of thermal fluctuation. The curve for $H \parallel ab$ peaks at 112 K, while the curve for $H \perp ab$ go on increase until 58 K. This behavior is consistent with above susceptibility measurements.

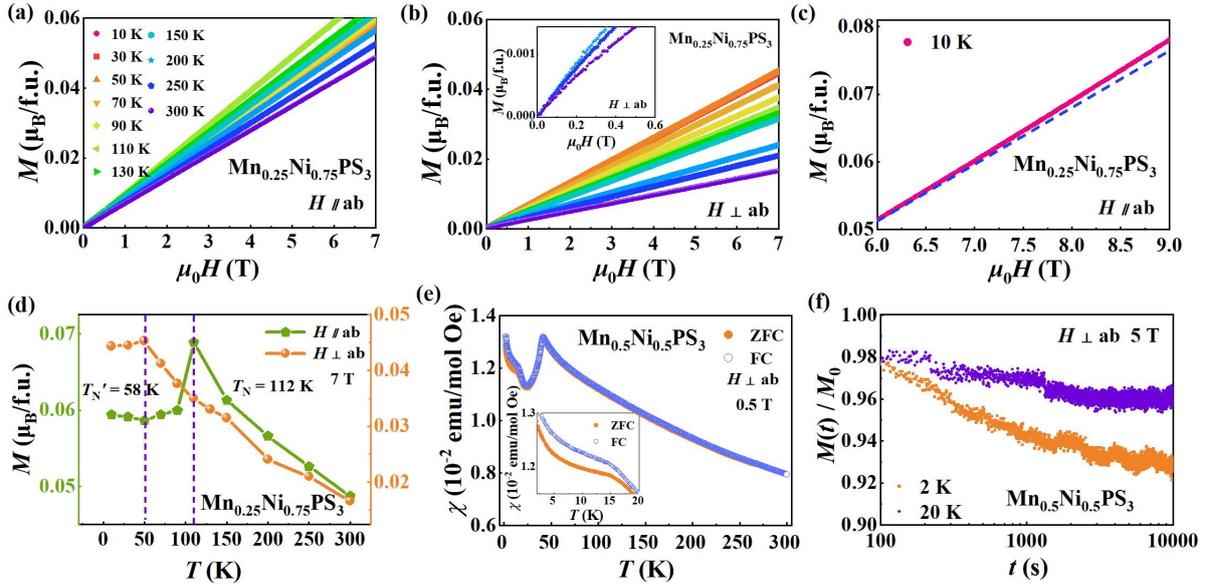

**Figure 4.** (a-b) Isothermal magnetization of $Mn_{0.25}Ni_{0.75}PS_3$ at different temperatures. (a) $H \parallel ab$. (b) $H \perp ab$. Inset: low field behavior. (c) Magnetization's deviation from linear behavior (blue dash line) under high field with $H \parallel ab$ at 10 K. (d) The magnetic moments at 7 T as a function of temperature under $H \parallel ab$ and $H \perp ab$ respectively. (e) Susceptibilities as a function of temperature for a field of 0.5 T for $Mn_{0.5}Ni_{0.5}PS_3$ with ZFC and FC sequences. Inset: susceptibilities between 2 K and 20 K. (b) Time dependence of thermoremanent magnetization for $Mn_{0.5}Ni_{0.5}PS_3$ at 2 K and 20 K respectively.

This nontrivial magnetic behavior can be understood as follows. The first possibility is that the in-plane component of spins is antiferromagnetically ordered at below 112 K, and the out-of-plane component becomes ordered below 58 K. This behavior may associate with the spin reorientation. Another possibility is that two magnetic phases with different magnetic easy axes are ordered at different temperatures, i.e. phase separation. To determine the real origin, more microscopic studies are needed, such as neutron scattering. In addition, for $H \parallel ab$ at 10 K, the *M-H* curve deviates from the linear dependence for field larger than 7 T, as shown in figure 4(c). In collinear AFM systems, the magnetic moment will rotate when external field along easy axis excess a critical field. A canting configuration and resulted net moments along the easy axis are formed. This is called spin-flop transition. The high field upturn of magnetization for $H \parallel ab$ are very weak and only observed at 10 K. A higher field magnetization measurement is desired to further clarify this issue. We also observed low-field sublinear behavior for *M-H* curve with $H \perp ab$ as shown in the inset of figure 4(b). This behavior was previously observed in Zn-doped MnPS$_3$ [49, 50]. It is attributed to the canted moment due to substitution. All of these signatures exhibit the complexity of magnetic structure for the 75% Ni substituted sample.

According to the temperature dependence of magnetic susceptibility, the ZFC and FC curves of entire series coincide completely except for the *x*=0.5 compound. As shown in figure 4(e), the magnetic susceptibility for $H \perp ab$ peaks at $T_N$=42 K. Irreversibility between ZFC and FC curves appear below 30 K, which may due to the appearance of spin glass. To verify this issue, the time dependence of thermoremanent magnetization is measured at 20 K and 2 K respectively, as shown in figure 4(f). It can be fitted by $M(t)/M(0) \propto \exp[-(t/\tau)]$, where $\tau \sim 700$ s at 2 K. Noting that similar exponential magnetic relaxation were also observed in Fe$_{0.5}$Ni$_{0.5}$PS$_3$, which was attributed to the magnetic glass [39]. In the FC measurement of Fe$_{0.5}$Ni$_{0.5}$PS$_3$, there was a thermal hysteresis between curves with increasing and decreasing temperature, but no irreversibility between the ZFC and FC measurement with increasing temperature. In contrast, in Mn$_{1-x}$Fe$_x$PS$_3$, the irreversibility between ZFC and FC curves was observed in a wide *x* range, and thus it was concluded as the spin glass state [38]. Our magnetization measurements on Mn$_{0.5}$Ni$_{0.5}$PS$_3$ imply a spin glassy like behavior.

To further understand above experimental results, a first principle DFT calculation was performed. First, we checked the bulks of two end members. Four most possible magnetic orders, including the ferromagnetic one and three antiferromagnetic ones are considered (see figure S2 in Supplementary Materials). Our DFT calculation confirms that the magnetic ground states are indeed the Néel type antiferromagnetism for MnPS$_3$ and zigzag type antiferromagnetism for NiPS$_3$, as summarized in Table I. And their optimized lattice constants are close to the experimental ones. These results provide a good start point to study the mixed cases.

Then the Ni substitution is tested, with the concentrations of 25%, 50%, and 75%, also summarized in Table I. For $x$=25%, it exhibits the Néel type antiferromagnetism, similar to MnPS$_3$. With 75% Ni substitution, it exhibits the zigzag type antiferromagnetism, in consistent with NiPS$_3$. While for the half-substituted case, two configurations are adopted (see figure S3(b-c) in Supplementary Materials), both of which lead to the zigzag type antiferromagnetism. Therefore, the magnetic transition from Néel type to zigzag type should occur between $x$=0.25 and $x$=0.5. Quantitatively, the energy differences between the ferromagnetic one and ground state one is reduced comparing with the two end members, especially for the half-substituted case, in consistent with the V-shape $T_N$ shown in figure 3(c).

**Table I.** DFT results of the magnetic ground state in Mn$_{1-x}$Ni$_x$PS$_3$. The experimental lattice constants of two end members are listed in parenthesis for comparison [51]. There are two configurations for Mn$_{0.5}$Ni$_{0.5}$PS$_3$ (see figure S3(b-c) in Supplementary Materials). The energies are in units of meV/f.u. and the ferromagnetic states are taken as the reference. The ground state ones are emphasized using bond fonts.

| System | | Energy (meV/f.u.) | | | Lattice constants | Gap (eV) |
|---|---|---|---|---|---|---|
| | | Néel | zigzag | stripy | ($a$, $b$, $c$) (Å) | |
| MnPS$_3$ | | **-41.8** | -24.5 | -24.4 | 6.071, 10.522, 6.812 (6.077, 10.524, 6.796) | 1.73 |
| NiPS$_3$ | | -43.8 | **-51.9** | 8.4 | 5.812, 10.068, 6.642 (5.812, 10.070, 6.632) | 1.08 |
| Mn$_{0.75}$Ni$_{0.25}$PS$_3$ | | **-35.3** | -28.1 | -12.2 | 6.048, 10.477, 6.805 | 0.82 |
| Mn$_{0.5}$Ni$_{0.5}$PS$_3$ | 1 | -33.4 | **-38.7** | 0.5 | 5.929, 10.293, 6.742 | 0.85 |
| | 2 | -29.4 | **-32.5** | 2.6 | 5.931, 10.297, 6.724 | 0.99 |

| | | | | | |
|---|---|---|---|---|---|
| Mn$_{0.25}$Ni$_{0.75}$PS$_3$ | -35.5 | **-42.2** | 5.4 | 5.869, 10.185, 6.693 | 0.88 |

## 4. Conclusion

We successfully grown a series of (Mn$_{1-x}$Ni$_x$)PS$_3$ single crystals by chemical vapor transport method. The systematic study on physical properties was performed by a combination of XRD, EDXs, DC magnetic susceptibility, thermoremanent magnetization measurements. For this series, the magnetism is expected to evolve with the Ni concentration, since their spins (moment, easy axis, anisotropy, interactions) are different between two end members. Some nontrivial magnetic properties have been found, with a V-shape curve of $T_N$. For the $x$=0.5 case, the irreversibility between ZFC and FC curves is observed below the magnetic ordering temperature for magnetic field applied perpendicular to *ab* plane, indicating a reentrant spin glass phase. For the $x$=0.75 case, abnormal bifurcation of $T_N$ is found. This magnetic evolution is further confirmed by DFT calculation. More characterizations such as neutron diffraction and magnetic force microscopy can be used to further reveal more microscopic details.

**Acknowledgements**

The work at Southeast University was supported by the National Natural Science Foundation of China (Grant nos. 11834002 and 52130706). Work at Nanjing University was supported by the National Natural Science Foundation of China (Grant nos. 11834002 and 51721001). Work at Hubei Normal University was supported by the National Natural Science Foundation of China (Grant no. 12074111).

Supplemental Materials of

# Evolution of magnetic phase in two dimensional van der Waals $Mn_{1-x}Ni_xPS_3$ single crystals

Ziye Lu[1], Xinyu Yang[1], Lin Huang[2], Xiyu Chen[3], Meifeng Liu[3], Jin Peng[1*], Shuai Dong[1*], and Jun-Ming Liu[2]

4. School of Physics, Southeast University, Nanjing 211189, China

5. Laboratory of Solid State Microstructures, Nanjing University, Nanjing 210093, China

6. Institute for Advanced Materials, Hubei Normal University, Huangshi 435002, China

*E-mails: jpeng@seu.edu.cn; sdong@seu.edu.cn

**Table S1.** The EDS atomic percents of $Mn_{1-x}Ni_xPS_3$ series.

| | Mn | Ni | P | S | Mn:Ni |
|---|---|---|---|---|---|
| $MnPS_3$ | 20.14 | 0 | 20.85 | 59.01 | 1:0 |
| $Mn_{0.75}Ni_{0.25}PS_3$ | 14.43 | 5.18 | 20.57 | 59.94 | 0.74:0.26 |
| $Mn_{0.5}Ni_{0.5}PS_3$ | 9.74 | 9.63 | 20.56 | 60.07 | 0.5:0.5 |
| $Mn_{0.25}Ni_{0.75}PS_3$ | 3.84 | 15.59 | 20.39 | 60.18 | 0.24:0.76 |
| $NiPS_3$ | 0 | 19.54 | 20.37 | 60.09 | 0:1 |

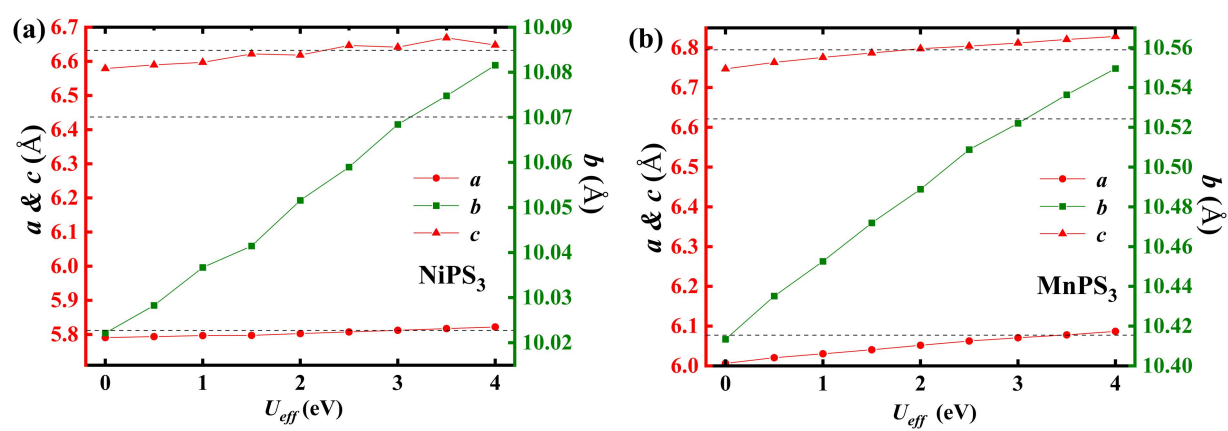

**Figure S1.** DFT testing of lattice constants of (a) $NiPS_3$ and (b) $MnPS_3$ as a function of $U_{eff}$ (eV). In both systems, $U_{eff}$ =3 eV seems to be the best choice, which leads to very tiny deviations from the experimental (*a*, *b*, *c*): (0.003%, 0.016%, 0.145%) for $NiPS_3$ and (0.1%, 0.02%, 0.23%) for (b) $MnPS_3$ respectively.

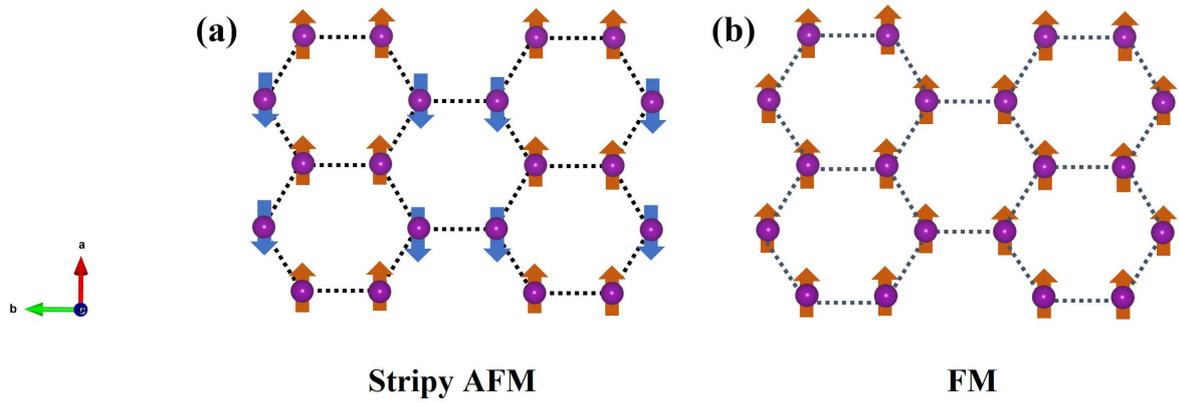

**Stripy AFM**  **FM**

**Figure S2.** The magnetic configurations used for energy comparison. (a) Stripy type antiferromagnetism. (b) Ferromagnetism. Other two configurations, i.e., Nèel type and zigzag type are shown in figure 1 of main text.

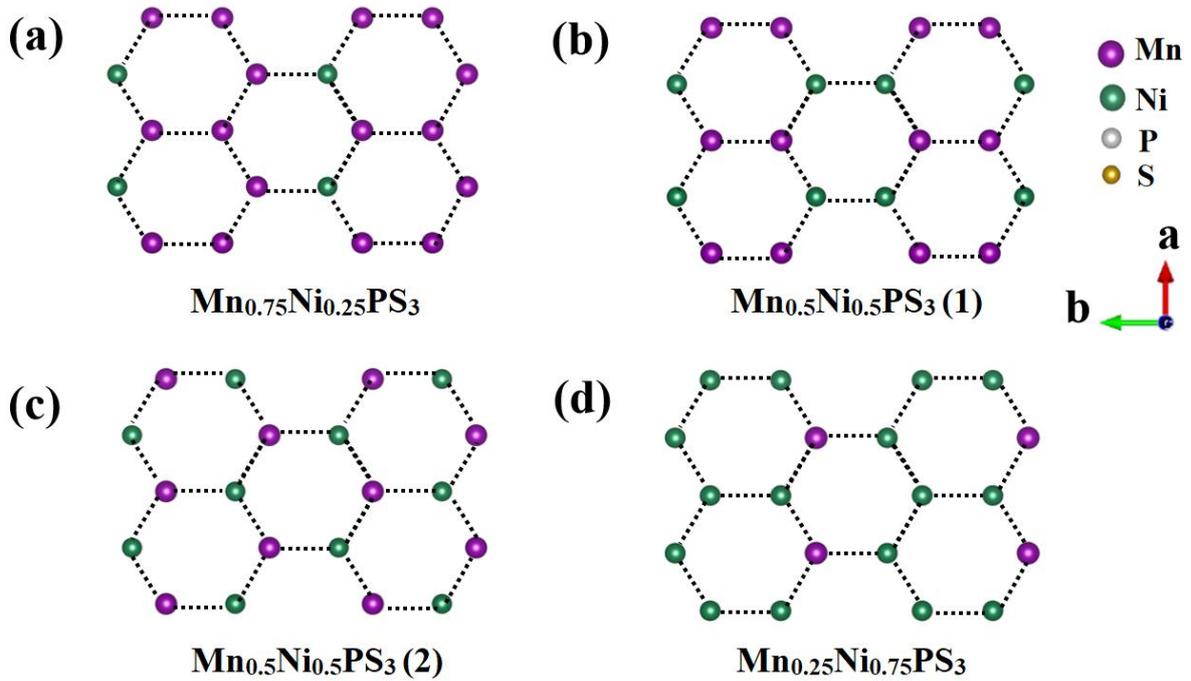

**Figure S3.** The substitution configurations of $Mn_{1-x}Ni_xPS_3$ considered in our DFT. (a) 25% Ni. (b-c) Two kinds of 50% Ni. (d) 75% Ni.